# Comments on "Bifurcations in DC-DC Switching Converters: Review of Methods and Applications"

Chung-Chieh Fang *

Manuscript: July 11, 2018

## Abstract

In a review paper [1] (El Aroudi, et al., 2005), two stability conditions for DC-DC converters are presented. However, these two conditions were published years earlier at least in a journal paper [2] (Fang and Abed, 2001). In this note, the similar texts of [1] and [2] are compared.

## 1 Introduction

A review paper is supposed to synthesize the primary sources on a particular topic, with those primary sources properly cited. However, many key references were not cited in the review paper [1], at least on two important closed-form stability conditions published years earlier for DC-DC converters. This note compares the similar texts in [1] and [2] (*as an example*).

## 2 Two Stability Conditions Presented in [2]

The first condition is summarized as follows. Consider a converter shown in Fig. 1. The linearized sampled-data dynamics is

$$\hat{x}_{n+1} \approx \Phi_o \hat{x}_n + \Gamma_v \hat{v}_{sn} + \Gamma_c \hat{v}_{cn} \tag{40}$$
$$\hat{v}_{on} = E\hat{x}_n \tag{41}$$

where the (Jacobian) matrix

$$\Phi_o = e^{A_2(T-d)}(I - \frac{(\dot{x}^0(d^-) - \dot{x}^0(d^+))F}{F\dot{x}^0(d^-) + \dot{h}(d)})e^{A_1 d} \tag{42}$$

The converter is stable if all of the eigenvalues of $\Phi_o$ are inside the unit circle of the complex plane.

Next, about the second (necessary) condition, the following is a direct quote from pp. 358-359 of [2].

"**Theorem 3:** *If the periodic solution $x^0(t)$ is open-loop asymptotically orbitally stable, then the following inequality holds*

$$\left| \frac{F\dot{x}^0(d^+) + \dot{h}(d)}{F\dot{x}^0(d^-) + \dot{h}(d)} \right| \leq e^{\text{tr}[A_2-A_1]d - \text{tr}[A_2]T} \tag{45}$$

*C.-C Fang is with Advanced Analog Technology, 2F, No. 17, Industry E. 2nd Rd., Hsinchu 300, Taiwan, Tel: +886-3-5633125 ext 3612, Email: fangcc3@yahoo.com

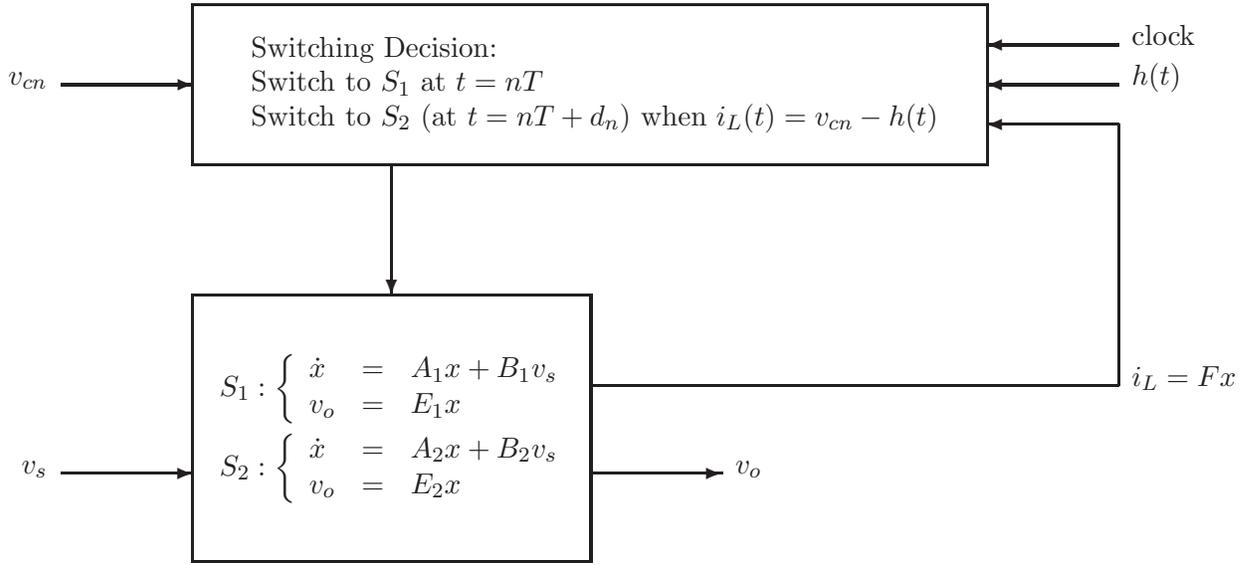

Figure 1: Power stage of PWM converter in CCM under current mode control

**Proof:** Suppose the periodic solution $x^0(t)$ is open-loop asymptotically orbitally stable. Then all the eigenvalues of $\Phi_o$ have magnitude less than or equal than 1. Since $\det[\Phi_o]$ is the product of the eigenvalues of $\Phi_o$, we have that

$$\begin{aligned}
|\det[\Phi_o]| &= \left|\det[e^{A_1 d}e^{A_2(T-d)}]\det[I - \frac{(\dot{x}^0(d^-) - \dot{x}^0(d^+))F}{F\dot{x}^0(d^-) + \dot{h}(d)}]\right| \\
&= \left|\det[e^{A_1 d}e^{A_2(T-d)}]\det[1 - \frac{F(\dot{x}^0(d^-) - \dot{x}^0(d^+))}{F\dot{x}^0(d^-) + \dot{h}(d)}]\right| \\
&= e^{-\text{tr}[A_2 - A_1]d + \text{tr}[A_2]T}\left|\frac{F\dot{x}^0(d^+) + \dot{h}(d)}{F\dot{x}^0(d^-) + \dot{h}(d)}\right| \\
&\leq 1
\end{aligned}$$

Then equation (45) follows. □

**Remark**: Generally the switching period is so small that the right side of (45) can be approximated as 1, resulting in a condition that resembles a well-known stability criterion in current mode control (Erickson (1997) for example):

$$\left|\frac{-m_2 + m_c}{m_1 + m_c}\right| < 1 \qquad (46)$$

where $m_1$, $m_2$, $m_c > 0$ and $m_1$ is the (positive) slope of the *inductor current* trajectory during the on stage and $-m_2$ is the (negative) slope during the off stage using a *linear approximation* (Middlebrook and Ćuk 1976); and $-m_c$ is the (negative) slope of the compensating ramp. The stability criterion (46) differs from Theorem 3, in which the *instantaneous* slope is used."

In [3] or [4, pp. 12-13], a similar necessary condition is also presented. The texts of [3, 4] are not presented here to save space.

## 3 Quoted Texts Related to the Two Stability Conditions From [1]

The following is a direct quote from p. 1560 of [1], about the first condition.

"It can be shown that the expression of the Jacobian matrix has the following form:

$$DP = \Phi_2(T^* - \tau^*)Q\Phi_1(\tau^*) \tag{32}$$

where Q is an appropiate matrix which depends on the controller used and $\tau^*$ and $T^*$ are, respectively, the switching instant and the duration of the cycle at the steady state. In the case of FFC, it can be demonstrated that

$$Q = I - \frac{(T^{*-} - \tau^{*+})K}{K\tau^{*-} - m_c} \text{ "} \tag{33}$$

The following is a direct quote from p. 1561 of [1], about the second condition.

"**6.1. A necessary condition for stability**
If the system is stable, both eigenvalues of the Jacobian matrix are inside the unit circle. The determinant of the Jacobian matrix which is equal to the product of its eigenvalues will be less than unity. Thus, for the case of FFC, by using Eq. (32), a necessary condition for local stability is:

$$\left|\frac{K\dot{x}^{*+} - m_c}{K\dot{x}^{*-} - m_c}\right| e^{-\text{tr}(A_2-A_1)d + \text{tr}(A_2)T} \leq 1 \tag{36}$$

It should be noted that this inequality is very similar to a well known and widely used criterion for stability [Erickson, 1997] which is:

$$\left|\frac{m_2 - m_c}{m_1 - m_c}\right| < 1 \tag{37}$$

where $m_1$ and $m_2$ are the positive and negative slopes of control signal, respectively and $m_c$ is the slope of the ramp compensator. However (36) is only a necessary and not sufficient condition for stability. Therefore traditional methods based in the approximated criterion (37) could fail to predict stability if the exponents in (36) are not sufficiently small."

It is interesting to note that by the time [1] was submitted in 2003, the book by Erickson already had a second edition in 2001. However, still the older 1997 edition (same as in [2]) was cited.

## 4 Discussion and Conclusion

The stability conditions such as (42) and (45) are important because they lead to many other important critical conditions of DC-DC converters [2, 3, 5-20]. The condition (42) was first known in [5, 6] in 1997-1998, and the condition (45) was first known in [3, 5] in 1997-1999. Before the submission of [1] in 2003, these two stability conditions had been already published years earlier in many references [2, 3, 5-12]. However, none of these earlier references are cited in [1]. Furthermore, the quoted text in [1] above has close similarity with the quoted text in [2]. For example, (32) and (36) of [1] are exactly the same as (42) and (45) of [2], respectively, just with different notations.